\documentclass[twocolumn]{jpsj3}

\usepackage{color}
\usepackage{times}
\usepackage{graphicx}

\title{Effect of $\Gamma_7$ and $\Gamma_8$ Hybridizations \\
on Three-Channel Kondo Phase Emerging from Ho Ions}

\author{Takashi Hotta}

\inst{Department of Physics, Tokyo Metropolitan University,
1-1 Minami-Osawa, Hachioji, Tokyo 192-0397, Japan}

\recdate{\today}

\abst{
By employing a numerical renormalization group method,
we analyze a seven-orbital impurity Anderson model
for Ho$^{3+}$ ion with ten $4f$ electrons.
This model includes both $V_7$ and $V_8$, which are hybridizations
between localized $4f$- and conduction electrons
in $\Gamma_7$ and $\Gamma_8$ orbitals, respectively.
For the case of $V_7=V_8$ with the local $\Gamma_5$ triplet ground state,
we have reported the discovery of a three-channel Kondo (TCK) phase,
characterized by a residual entropy of $\log \phi$
with the golden ratio $\phi=(1+\sqrt{5})/2$.
In this research, by depicting the ground-state phase diagram
on the $(V_8, V_7)$ plane,
we attempt to unveil the effect of $V_7$ and $V_8$
on the emergence of the TCK phase.
After performing a lot of numerical calculations,
we find that the TCK phase appears in a relatively wide region
on the $(V_8, V_7)$ plane.
The boundary curves surrounding the TCK phase are determined
by the variation of the temperature dependence in entropy
and the abrupt change in energy spectra.
We consider that most of the phases surrounding the TCK phase
are Fermi liquids, but the non-Fermi liquid two-channel Kondo
phase is unexpectedly found to exist next to the TCK phase.
Finally, we briefly comment on the actual material
concerning the detection of the TCK phase.
}

\begin{document}
\maketitle

\section{Introduction}

More than four decades ago, Nozi\`eres and Blandin have
proposed a novel concept of two-channel Kondo effect,
\cite{Nozieres}
originating from the overscreening of impurity spin $S=1/2$
by two-channel conduction electron bands.
This exciting proposal has opened a new door,
leading to a potential source of exotic quantum ground states
such as a non-Fermi liquid phase.
After the proposal of the idea of the two-channel Kondo effect,
it has been immediately extended to the concept of
multi-channel Kondo effect, but in any case,
theoretical research has been preceded first.

However, the situation has been drastically changed,
when Cox has pointed out the existence of two screening channels
in terms of quadrupole degrees of freedom in U$^{4+}$ ($5f^2$)
systems with non-Kramers doublet ground state.\cite{Cox1,Cox2}
Then, experimental studies to observe the two-channel Kondo effect
have had significant advances in observing the signals of
the two-channel Kondo effect in cubic uranium compounds
with non-Kramers doublet ground state.
In the present century, the main target for the two-channel Kondo
effect has moved from U$^{4+}$ to Pr$^{3+}$ ($4f^2$) systems
and the signals of the two-channel Kondo effect have been actually
observed.\cite{Sakai,Onimaru1,Onimaru2,Higashinaka,Review}

The quadrupole Kondo phenomenon has been considered to
be the central issue to realize the two-channel Kondo effect,
but it is believed to be important to expand the research
frontier of multi-channel Kondo physics
in rare-earth and actinide ions other than Pr$^{3+}$ and U$^{4+}$.
In this viewpoint, it has been shown that the two-channel Kondo effect
emerges in Nd$^{3+}$ ($4f^3$) for a wide range of parameters
with the local $\Gamma_6$ doublet ground state.\cite{Hotta-Nd}
This is considered to be the magnetic two-channel Kondo effect,
when we recall the original concept by Nozi\`eres and Blandin.
A possibility of the occurrence of the two-channel Kondo effect
in $5f^4$ systems such as Np$^{3+}$ and Pu$^{4+}$ ions
has been also pointed out.\cite{Matsui-Hotta}

In addition to the discovery of new stages for the two-channel Kondo effect,
it is also interesting to pursue the realization of the multi-channel Kondo
phenomena beyond the two-channel Kondo effect.
As for this point, in a three-orbital impurity Anderson model
for a single C$_{60}$ molecule, Leo and Fabrizio have discussed
the phase diagram including the three-channel Kondo state.\cite{Fabrizio}
By analyzing a seven-orbital impurity Anderson model
hybridized with $\Gamma_7$ and $\Gamma_8$ conduction electrons
for Ho$^{3+}$ ions with ten $4f$ electrons,
the present author has discovered the three-channel Kondo effect 
for the local $\Gamma_5$ triplet ground state,\cite{Hotta-Ho}
characterized by a residual entropy of $\log \phi$ with the golden ratio
$\phi=(1+\sqrt{5})/2$.

In this study, we attempt to deepen our understanding
on the emergence of the three-channel Kondo effect from Ho ions
for the case with the local $\Gamma_5$ triplet ground state.
For the purpose, we investigate a quantum critical point (QCP)
around the three-channel Kondo state in the phase diagram
on the ($V_8$, $V_7$) plane,
where $V_8$ and $V_7$ denote the hybridization of
localized $\Gamma_8$ and $\Gamma_7$ electrons
with the conduction bands, respectively.
In the previous paper,\cite{Hotta-Ho}
we have considered only the case of $V_7=V_8$,
but here we depict the phase diagram on the ($V_8$, $V_7$) plane
to unveil how the three-channel Kondo phase emerges
from the QCP's in the phase diagram.

The paper is organized as follows.
In Sect.~2, we explain the local model including spin-orbit coupling,
crystalline electric field (CEF) potentials,
and Coulomb interactions among $f$ electrons.
Then, we construct a seven-orbital impurity Anderson model by including
further the hybridization between localized and conduction electrons
in $\Gamma_7$ and $\Gamma_8$ orbitals.
We also briefly explain a numerical renormalization group (NRG) method
to analyze the model Hamiltonian.
In Sect.~3, first we briefly review the previous results
on the three-channel Kondo effect for the case of $V_7=V_8$.
Next we show the present results for the general case of $V_7 \ne V_8$
to depict the ground-state phase diagram on the $(V_8, V_7)$ plane.
We explain the determination of the boundary curves in the phase diagram
by the entropy behavior and the changes in the energy spectra.
Finally, in Sect.~4, we summarize this paper and
provide a few comments on the future problems.
We also briefly comment on the detection of the three-channel Kondo
effect in actual materials.
Throughout this paper, we use such units as $\hbar=k_{\rm B}=1$
and the energy unit is set as eV.

\section{Model and Method}

In this section, we explain the construction of a seven-orbital
impurity Anderson model.
Note that the model Hamiltonian itself has been already shown
in the previous papers,
but to make this paper self-contained, we improve the explanation
to construct the model Hamiltonian in this opportunity.
In particular, we explain the description of the local $f$-electron state
on the basis of a $j$-$j$ coupling scheme.

\subsection{Local $f$-electron model}

Let us start our explanation on the definition of the local $f$-electron
Hamiltonian $H_{\rm loc}$, composed of a spin-orbit coupling,
CEF potentials, and Coulomb interaction terms.
We express $H_{\rm loc}$ as
\begin{equation}
\label{Hloc1}
\begin{split}
    H_{\rm loc} &= \sum_{m,\sigma,m',\sigma'}
    (\zeta_{m,\sigma;m',\sigma'}\!+\! \delta_{\sigma,\sigma'} B_{m,m'})
    f_{m\sigma}^{\dag}f_{m'\sigma'} \\
   &+\sum_{m_1 \sim m_4}\sum_{\sigma,\sigma'} I_{m_1m_2,m_3m_4}
    f_{m_1\sigma}^{\dag}f_{m_2\sigma'}^{\dag}f_{m_3\sigma'}f_{m_4\sigma}\\
   &+ n E_{f},
\end{split}
\end{equation}
where $f_{m\sigma}$ denotes an annihilation operator
for local $f$ electron with spin $\sigma$ and
$z$-component $m$ of angular momentum $\ell=3$,
$\sigma=\uparrow$ ($\downarrow$) for up (down) spin,
$\zeta$ is the matrix element for the spin-orbit coupling,
$B_{m,m'}$ indicates CEF potentials for $f$ electrons
from the ligand ions,
$I$ is the matrix element of Coulomb interactions,
$n$ is the local $f$-electron number at an impurity site,
and $E_f$ is the $f$-electron level to control $n$.
Note that $\sigma$ is also defined as a variable to
take $\sigma=+1$ and $-1$ for up and down spin, respectively.

Concerning the matrix element for the spin-orbit coupling,
$\zeta$ is explicitly written as
\begin{equation}
\begin{split}
    \zeta_{m,\sigma;m,\sigma}&
       =\frac{\lambda m\sigma}{2},\\
    \zeta_{m+\sigma,-\sigma;m,\sigma}&
      =\frac{\lambda \sqrt{\ell(\ell+1)-m(m+\sigma)}}{2},
\end{split}
\end{equation}
and zeros for other cases,
where $\lambda$ is a spin-orbit coupling constant.
In this paper, we set $\lambda=0.265$ eV for Ho ion.
\cite{spin-orbit}

As for the CEF potentials,
$B_{m,m'}$ is defined in the table of Hutchings
for the angular momentum $\ell=3$.\cite{Hutchings}
For cubic structure with $O_{\rm h}$ symmetry,
$B_{m,m'}$ is given by the fourth- and sixth-order
CEF potential parameters, $B_4^0$ and $B_6^0$, as
\begin{equation}
\begin{split}
    B_{3,3}&=B_{-3,-3}=180B_4^0+180B_6^0, \\
    B_{2,2}&=B_{-2,-2}=-420B_4^0-1080B_6^0, \\
    B_{1,1}&=B_{-1,-1}=60B_4^0+2700B_6^0, \\
    B_{0,0}&=360B_4^0-3600B_6^0, \\
    B_{3,-1}&=B_{-3,1}=60\sqrt{15}(B_4^0-21B_6^0),\\
    B_{2,-2}&=300B_4^0+7560B_6^0.
\end{split}
\end{equation}
Here we note the relation of $B_{m,m'}=B_{m',m}$.
Following the traditional notation in Ref.~[\citen{LLW}],
we redefine $B_4^0$ and $B_6^0$ as
\begin{equation}
    B_4^0=\frac{Wx}{F(4)},~B_6^0=\frac{W(1-|x|)}{F(6)},
\end{equation}
where $x$ specifies the CEF scheme for the $O_{\rm h}$ point group,
while $W$ determines the energy scale for the CEF potential.
We choose $F(4)=15$ and $F(6)=180$ for $\ell=3$.\cite{Hutchings}
In this paper, we set $W=10^{-3}$ eV and treat $x$ as a parameter
to control the CEF ground state between $-1 \le x \le 1$.

Finally, the matrix element of Coulomb interactions $I$ is given by
\begin{equation}
I_{m_1m_2,m_3m_4} = \sum_{k=0}^{6} F^k c_k(m_1,m_4)c_k(m_2,m_3).
\end{equation}
Here $F^k$ indicates the Slater-Condon parameter and
$c_k$ is the Gaunt coefficient.\cite{Slater}
Note that the sum is limited by the Wigner-Eckart theorem to
$k=0$, $2$, $4$, and $6$.
Although the Slater-Condon parameters should be determined
for the material from the experimental results,
here we set the ratio as
\begin{equation}
  \frac{F^0}{10}=\frac{F^2}{5}=\frac{F^4}{3}=F^6=U,
\end{equation}
where $U$ indicates the Hund's rule interaction among $f$ orbitals.
In this paper, we set $U=1$ eV.

\subsection{Local model on the basis of a $j$-$j$ coupling scheme}

It is not difficult to obtain the local $f$ electron states
by performing the exact diagonalization of $H_{\rm loc}$,
but it is more convenient to change the $f$-electron bases
for the construction of the impurity Anderson model.
\cite{Hotta-Ueda,Hotta-Harima}
First we define the one-electron states by the cubic irreducible
representations.
Then, we include Coulomb interactions among $f$ electrons.

For the purpose to diagonalize the spin-orbit coupling term,
we transform the $f$-electron basis
between $(m,\sigma)$ and $(j,\mu)$ representations,
connected by Clebsch-Gordan coefficients,
where $j$ is the total angular momentum and
$\mu$ is the $z$-component of $j$.
Hereafter we use symbols ``$a$'' and ``$b$'' for $j=5/2$ and $7/2$,
respectively.
When we define $f_{j\mu}$ as the annihilation operator
for $f$ electron labeled by $j$ and $\mu$,
the transformation is expressed as
\begin{equation}
   f_{j\mu} = \sum_{m,\sigma} C^{(j)}_{\mu;m,\sigma} f_{m\sigma},
\end{equation}
where the Clebsch-Gordan coefficient $C^{(j)}_{\mu;m,\sigma}$ is given by
\begin{equation}
\begin{split}
   C^{(a)}_{\mu;\mu-\sigma/2,\sigma} &=
   -\sigma \sqrt{\frac{7/2-\sigma \mu}{7}},\\
   C^{(b)}_{\mu;\mu-\sigma/2,\sigma} &=
               \sqrt{\frac{7/2+\sigma \mu}{7}},
\end{split}
\end{equation}
and other components are zeros.

Next we introduce new operators characterized by
the cubic irreducible representation.
For the purpose, we diagonalize each CEF potential term
of $j=5/2$ and $7/2$ with the cubic symmetry.
After some algebraic calculations,
we obtain $\Gamma_7$ doublet and $\Gamma_8$ quartet
from $j=5/2$ sextet,
whereas $\Gamma_6$ doublet, $\Gamma_7$ doublet,
and $\Gamma_8$ quartet from $j=7/2$ octet.
Then, we define new operators with orbital degrees of freedom $\nu$
and pseudo-spin $\tau$ as
\begin{equation}
  f_{j,\nu,\tau}=\sum_{\mu} D^{(j)}_{\nu,\tau; \mu} f_{j\mu},
\end{equation}
where $\nu$ is the label to express the cubic irreducible representation,
$\tau=\uparrow$ ($\downarrow$) for up (down) pseudo-spin
to distinguish the Kramers doublet for each orbital,
and $D^{(j)}$ is the coefficient to connect
the $f$-electron base between $(j, \mu)$ and $(j,\nu,\tau)$.

For $j=a$ ($j=5/2$), we define $\nu=\alpha$ and $\beta$ for
$\Gamma_8$ quartet, while $\nu=\gamma$ is introduced for
$\Gamma_7$ doublet.
Explicitly, we obtain $D^{(a)}$ as
\begin{equation}
\begin{split}
D^{(a)}_{\alpha,\uparrow;-\frac{5}{2}}\!&
 =\!D^{(a)}_{\alpha,\downarrow;\frac{5}{2}}
\!=\! -D^{(a)}_{\gamma,\uparrow;\frac{3}{2}}\!
 =\!-D^{(a)}_{\gamma,\downarrow;-\frac{3}{2}}\!=\!\sqrt{\frac{5}{6}},\\
D^{(a)}_{\beta,\uparrow;-\frac{1}{2}}\!&
 =\!D^{(a)}_{\beta,\downarrow;\frac{1}{2}}=1,\\
D^{(a)}_{\alpha,\uparrow;\frac{3}{2}}\!&
 =\! D^{(a)}_{\alpha,\downarrow;-\frac{3}{2}}\!=
\! D^{(a)}_{\gamma,\uparrow;-\frac{5}{2}} \!
 =\! D^{(a)}_{\gamma,\downarrow;\frac{5}{2}}\!=\!\sqrt{\frac{1}{6}}.
\end{split}
\end{equation}
On the other hand, for $j=b$ ($j=7/2$), we define $\nu=\alpha$ and $\beta$ for
$\Gamma_8$ quartet, $\nu=\gamma$ for $\Gamma_7$ doublet,
and $\nu=\delta$ for $\Gamma_6$ doublet.
Then, we write $D^{(b)}$ as
\begin{equation}
\begin{split}
 D^{(b)}_{\alpha,\uparrow;-\frac{5}{2}}\!&
=\!-D^{(b)}_{\alpha,\downarrow;\frac{5}{2}}\!
=\!-D^{(b)}_{\gamma,\uparrow;\frac{3}{2}}\!
=\!D^{(b)}_{\gamma,\downarrow;-\frac{3}{2}}\!
=\!\frac{1}{2},\\
 D^{(b)}_{\alpha,\uparrow;\frac{3}{2}}\!&
=\!-D^{(b)}_{\alpha,\downarrow;-\frac{3}{2}}\!
=\!D^{(b)}_{\gamma,\uparrow;-\frac{5}{2}}\!
=\!-D^{(b)}_{\gamma,\downarrow;\frac{5}{2}}\!
=\!\frac{\sqrt{3}}{2},\\
 D^{(b)}_{\beta,\uparrow;-\frac{1}{2}}\!&
=\!-D^{(b)}_{\beta,\downarrow;\frac{1}{2}}\!
=\! D^{(b)}_{\delta,\uparrow;\frac{7}{2}}\!
=\!-D^{(b)}_{\delta,\downarrow;-\frac{7}{2}}\!
=\!\sqrt{\frac{5}{12}},\\
 -D^{(b)}_{\beta,\uparrow;\frac{7}{2}}\!&
=\!D^{(b)}_{\beta,\downarrow;-\frac{7}{2}}\!
=\!D^{(b)}_{\delta,\uparrow;-\frac{1}{2}}\!
=\!-D^{(b)}_{\delta,\downarrow;\frac{1}{2}}\!
=\!\sqrt{\frac{7}{12}}.
\end{split}
\end{equation}
For the standard time reversal operator
${\cal K}=-{\rm i}\sigma_y K$,
where $K$ denotes an operator to take the complex conjugate,
we can easily show the relation~\cite{Hotta-Ueda}
\begin{equation}
 {\cal K}f_{j,\nu,\tau}=\tau f_{j,\nu,-\tau},
\end{equation}
where $\tau=+1$ ($-1$) for up (down) pseudo-spin.
Note that this has the same definition for real spin.

By using the new operator $f_{j,\nu,\tau}$,
we write the new local Hamiltonian,
composed of the seven orbitals characterized by
the cubic irreducible representation.
Then, the new local Hamiltonian is expressed as
\begin{equation}
\begin{split}
H_{\rm loc} &=\sum_{j, j', \nu, \tau} (\lambda_j \delta_{j,j'}+
B_{j,j',\nu})f_{j \nu \tau}^{\dag} f_{j' \nu \tau} + n E_{f} \\
&+\sum_{j_1\sim  j_4} \sum_{\nu_1 \sim \nu_4}
\sum_{\tau_1 \sim \tau_4} 
{\tilde I}^{j_1 j_2, j_3 j_4}_
{\nu_1 \tau_1 \nu_2 \tau_2, \nu_3 \tau_3 \nu_4 \tau_4} \\
&\times  f_{j_1 \nu_1 \tau_1}^{\dag} f_{j_2 \nu_2 \tau_2}^{\dag}
f_{j_3 \nu_3 \tau_3}  f_{j_4 \nu_4 \tau_4},
\end{split}
\end{equation}
where $\lambda_j$ is given by
\begin{equation}
   \lambda_a=-2\lambda,~\lambda_b=\frac{3}{2}\lambda.
\end{equation}
Concerning the CEF potential term,
the diagonal and off-diagonal parts are, respectively, given by
\begin{equation}
\begin{split}
   B_{a,a,\alpha} & =B_{a,a,\beta}=\frac{1320}{7}B_4^0,\\
   B_{a,a,\gamma} &=-\frac{2640}{7}B_4^0,\\
   B_{b,b,\alpha} &=B_{b,b,\beta}=\frac{360}{7}B_4^0+2880B_6^0,\\
   B_{b,b,\gamma} &=-\frac{3240}{7}B_4^0-2160 B_6^0,\\
   B_{b,b,\delta} &=360 B_4^0-\frac{3600}{7}B_6^0,
\end{split}
\end{equation}
and
\begin{equation}
\label{eq:CEFoff}
\begin{split}
   B_{a,b,\alpha} & =-B_{a,b,\beta}=-\frac{720}{7}\sqrt{5}B_4^0
                              +2160\sqrt{5}B_6^0,\\
   B_{a,b,\gamma} &=-\frac{1200}{7}\sqrt{3}B_4^0-4320\sqrt{3}B_6^0.
\end{split}
\end{equation}
Note the relation of $B_{j,j',\nu}=B_{j',j,\nu}$.

Concerning the CEF potential terms, three comments are in order.
First we emphasize that the off-diagonal CEF terms should appear
in the same orbital $\nu$ between $j=5/2$ and $7/2$.\cite{footnote}
Second we note that the CEF potentials are independent of pseudo-spin,
since they work only on the charge distribution.
Finally, we also note that $B_6^0$ does not appear for $j=5/2$,
since the maximum size of the change of the total angular momentum,
$2j=5$ in this case, is less than $2\ell=6$.

The Coulomb interaction ${\tilde I}$ is expressed as
\begin{equation}
\begin{split}
&{\tilde I}^{j_1 j_2, j_3 j_4}
_{\nu_1 \tau_1 \nu_2 \tau_2, \nu_3 \tau_3 \nu_4 \tau_4}
= \sum_{m_1\sim  m_4} \sum_{\sigma, \sigma'}
A^{(j_1)}_{\nu_1 \tau_1,m_1 \sigma} \\
&\times A^{(j_2)}_{\nu_2 \tau_2,m_2 \sigma'}
A^{(j_3)}_{\nu_3 \tau_3,m_3 \sigma'} A^{(j_4)}_{\nu_4 \tau_4,m_4 \sigma}
I_{m_1m_2,m_3m_4}
\end{split}
\end{equation}
where the coefficient $A$ is given by
\begin{equation}
A^{(j)}_{\nu,\tau,m,\sigma}=\sum_{\mu} D^{(j)}_{\nu,\tau; \mu}
C^{(j)}_{\mu;m,\sigma}.
\end{equation}

\begin{figure}[t]
\centering
\includegraphics[width=8truecm]{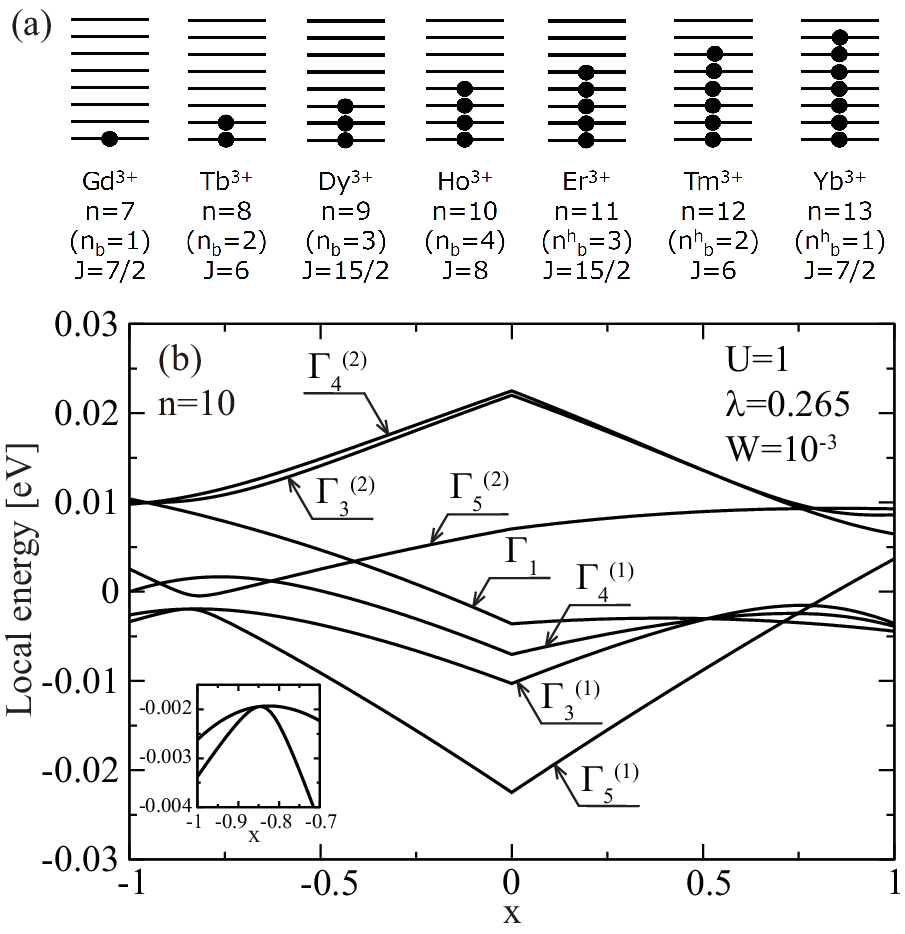}
\caption{
(a) Electron configurations in the $j=7/2$ octet for $7 \le n \le 13$.
Here we show $f$ electrons by solid circles.
Note that we omit the $j=5/2$ sextet which is fully occupied.
(b) Local energies vs. $x$ for $n=10$
with $U=1$, $\lambda=0.265$, and $W=10^{-3}$.
Inset shows the ground and first excited-state energies
for $-1.0 \leq x \leq -0.7$,
suggesting that $\Gamma_5$ triplet becomes the ground state
except for a very narrow region around at $x=-0.85$.
}
\end{figure}

Before proceeding to the introduction of
a seven-orbital impurity Anderson model,
we explain the specificity of Ho$^{3+}$ among rare-earth ions
based on the $j$-$j$ coupling scheme, as shown in Fig.~1(a).
We define $n_a$ and $n_b$ as $f$-electron numbers
in the $j=5/2$ sextet and $j=7/2$ octet, respectively.
We also define $n_b^{\rm h}=8-n_b$ as hole numbers
in the $j=7/2$ octet.
In analogy with the cases of $n_a=2$ (Pr$^{3+}$) and
$n_a=3$ (Nd$^{3+}$),
we expect the emergence of the two-channel Kondo effect
for $n_b=2$ (Tb$^{3+}$), $n_b=3$ (Dy$^{3+}$),
$n_b^{\rm h}=3$ (Er$^{3+}$),
and $n_b^{\rm h}=2$ (Tm$^{3+}$).
However, we accommodate four electrons in the $j=7/2$
octet for $n_b=4$ (Ho$^{3+}$), 
leading to the unique situation among rare-earth ions.
Then, we consider the case of $n=10$
to seek for the three-channel Kondo effect.

Next we briefly discuss the local ground states for $n=10$.
Without the CEF potentials, the ground-state multiplet for $n=10$
is characterized by the total angular momentum $J=8$.
When we apply the cubic CEF potentials, we notice that
the sept-dectet of $J=8$ is split into four groups
as one $\Gamma_1$ singlet, two $\Gamma_3$ doublets,
two $\Gamma_4$ triplets, and two $\Gamma_5$ triplets.
In Fig.~1(b), we depict the local energies as functions of $x$
for $W=10^{-3}$ by following the traditional manner.\cite{LLW}
As mentioned above, we actually observe one $\Gamma_1$ singlet,
two $\Gamma_3$ doublets, two $\Gamma_4$ triplets,
and two $\Gamma_5$ triplets.
Here we note that $W$ is defined as a positive value.
If we set $W<0$, the order in the eigenstates is reversed.
Namely, the $\Gamma_4^{(2)}$ triplet becomes the
ground state, whereas the $\Gamma_3^{(2)}$ doublet is
the first excited-state with a tiny excitation energy.
\cite{Hotta2005,Hotta2007}

Let us here focus on the ground state for the case of $W=10^{-3}$.
Roughly speaking, $\Gamma_5$ triplet ground state appears
widely for $-1 \leq x \leq 0.71$,
whereas $\Gamma_1$ singlet ground state appears for $0.71 \leq x \leq 1.0$.
As shown in the inset, $\Gamma_3$ doublet ground state appears 
only for a very narrow region around at $x=-0.85$,
but the quasi quintet is found to appear in the region of $-1.0 \leq x \leq -0.8$.

\subsection{Seven-orbital impurity Anderson model}

Now we construct a seven-orbital impurity Anderson model
by including the $\Gamma_7$ and $\Gamma_8$ conduction
electron bands hybridized with localized $f$ electrons.
Since here we discuss the case of $n=10$ (Ho$^{3+}$ ion),
the $j=5/2$ sextet is considered to be fully occupied and
the Fermi level should be situated among the $j=7/2$ octet.
Namely, it is necessary to take into account the hybridization
between the conduction and $j=7/2$ electrons in the present
research.

Then, the seven-orbital Anderson model is given by
\begin{equation}
H \!=\! \sum_{\mib{k},\nu,\tau} \varepsilon_{\mib{k}}
c_{\mib{k}\nu\tau}^{\dag} c_{\mib{k}\nu\tau}
\!+\! \sum_{\mib{k},\nu,\tau} V_{\nu}
(c_{\mib{k}\nu\tau}^{\dag}f_{b\nu\tau} \!+\! {\rm h.c.})
\!+\! H_{\rm loc},
\end{equation}
where $\varepsilon_{\mib{k}}$ is the dispersion of
the conduction electron with the wave vector $\mib{k}$,
$c_{\mib{k}\nu\tau}$ is the annihilation operator of the
conduction electron with orbital $\nu$ and pseudo-spin $\tau$,
and $V_{\nu}$ denotes the hybridization between
the localized and conduction electrons of the $\nu$ orbital.

In the previous paper, we have considered only the case of
$V_{\alpha}=V_{\beta}=V_{\gamma}=V$.\cite{Hotta-Ho}
As mentioned before, $V_{\alpha}$ should be equal to
$V_{\beta}$ from the cubic symmetry,
but $V_{\gamma}$ can take a different value from
$V_{\alpha}$ and $V_{\beta}$.
Thus, in this study, we define
\begin{equation}
V_{\alpha}=V_{\beta}=V_8,~~V_{\gamma}=V_7,
\end{equation}
and we will consider the general case of $V_8 \ne V_7$.

\subsection{Numerical renormalization group (NRG) method}

In this research, we analyze the seven-orbital impurity Anderson
model by using the NRG method,\cite{NRG1,NRG2}
in which we logarithmically discretize the momentum space
so as to include efficiently conduction electron states
near the Fermi energy.
Then, we characterize the conduction electron states
by shells labeled by $N$, and the shell of $N=0$ denotes
an impurity site described by $H_{\rm loc}$.
The NRG method has been explained in previous papers,
but to make this paper self-contained,
here we will briefly review the method.

After some algebraic calculations,
we can transform the Hamiltonian into a recursive form as
\begin{equation}
H_{N+1} = \sqrt{\Lambda} H_N + t_N \sum_{\nu,\tau}
(c_{N \nu \tau}^{\dag}c_{N+1 \nu \tau}+{\rm h.c.}),
\end{equation}
where $\Lambda$ denotes a parameter to control
the logarithmic discretization,
$c_{N \nu\tau}$ indicates the annihilation operator of
the conduction electron in the $N$-shell,
and $t_N$ is the ``hopping'' of the electron between
$N$- and $(N+1)$-shells, expressed by
\begin{equation}
t_N=\frac{(1+\Lambda^{-1})(1-\Lambda^{-N-1})}
{2\sqrt{(1-\Lambda^{-2N-1})(1-\Lambda^{-2N-3})}}.
\end{equation}
The initial term $H_0$ is given by
\begin{equation}
H_0=\Lambda^{-1/2} \left[ H_{\rm loc}
+\sum_{\nu,\tau} \left(c_{0 \nu \tau}^{\dag} f_{\nu \tau}
+{\rm h.c.} \right) \right].
\end{equation}

To calculate thermodynamic quantities,
we evaluate the free energy $F$ for the local $f$ electron in each step as
\begin{equation}
F_N = -T \left( \ln {\rm Tr} e^{-H_N/T}
- \ln {\rm Tr} e^{-H_N^0/T} \right),
\end{equation}
where $F_N$ denotes the free energy at the step $N$,
a temperature $T$ is defined as $T=\Lambda^{-(N-1)/2}$
at each step in the NRG calculation,
and $H_N^0$ indicates the free-electron part,
i.e., the Hamiltonian without the impurity and hybridization terms.
Then, we obtain the entropy $S_{\rm imp}$ as
$S_{\rm imp}=-\partial F/\partial T$.

In the NRG calculation, we keep $M$ low-energy states
in each renormalization step and
$M$ is set as $5,000$ in this research.
As for the value of $\Lambda$, we set $\Lambda=8.0$.
In the present NRG calculation, mainly to save of the CPU time,
we terminate the iteration at $N=30$.
Namely, the lowest temperature at which we arrive is
$T=8.0 \times 10^{-14}$.
Finally, the energy unit of the NRG calculation is
a half of conduction band width,
which is set as 1eV in the present research.

\begin{figure}[t]
\centering
\includegraphics[width=8truecm]{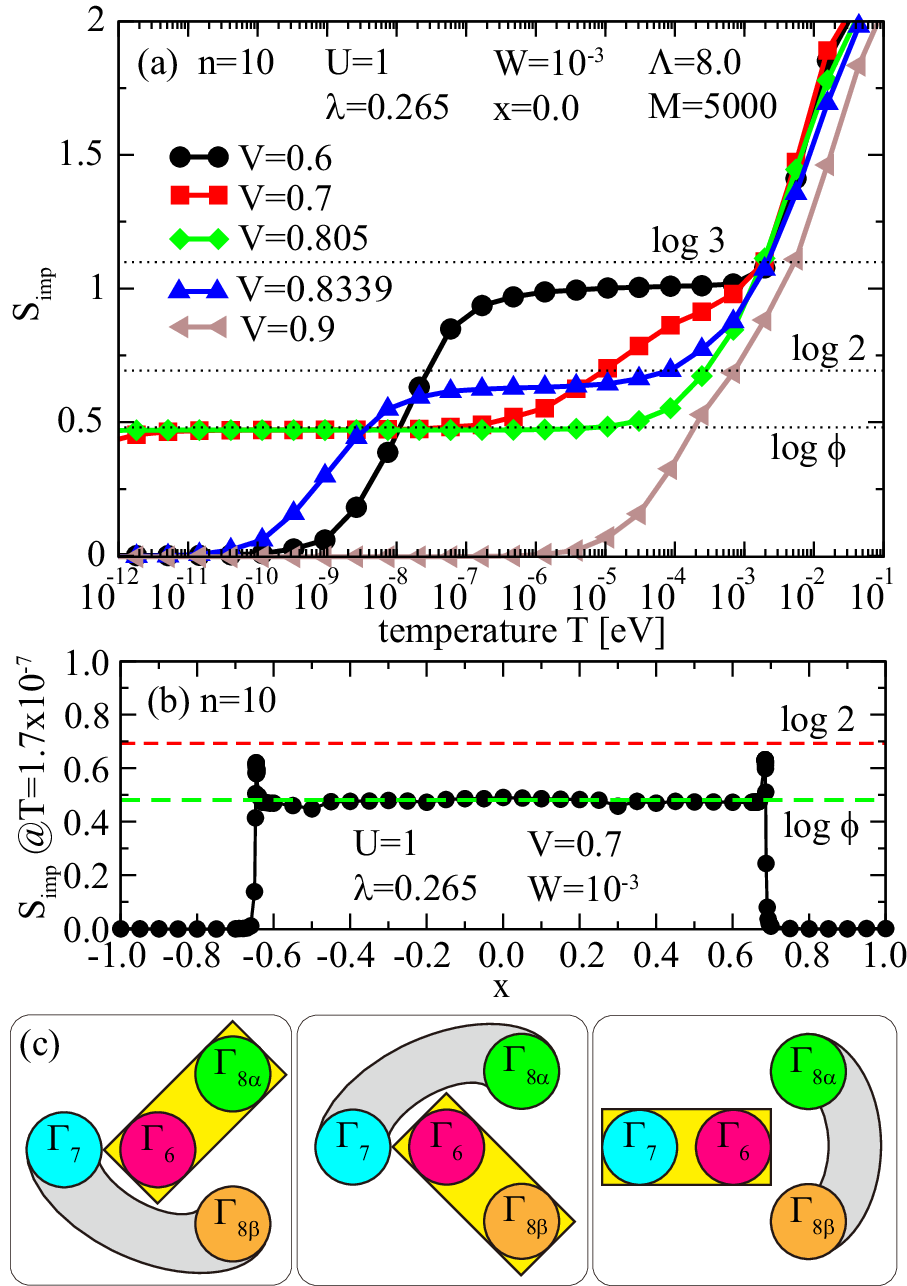}
\caption{(Color online)
(a) Entropies vs. temperature for several values of $V$
for $U=1$, $\Lambda=0.265$, $W=10^{-3}$, and $x=0.0$.
(b) Residual entropies at $T=1.7 \times 10^{-7}$ vs. $x$
for $V=0.7$ with $U=1$, $\Lambda=0.265$, and $W=10^{-3}$.
(c) Schematic views for the main components of the $\Gamma_5$
triplet of $n=10$. The rectangle and arc denote the triplet and
singlet pairs, respectively.\cite{Hotta-Ho}
}
\end{figure}

\section{Calculation Results}

\subsection{Review of the results for the case of $V_7=V_8$}

Before proceeding to the exhibition of the present results
for the general case of $V_7 \ne V_8$,
let us briefly review the previous results for the case of $V_7=V_8=V$.
In Fig.~2, we summarize the results of Ref.~\citen{Hotta-Ho}.
First we pay our attention to Fig.~2(a),
in which we show the NRG results of $f$-electron entropy
for $W=10^{-3}$ and $x=0.0$ with the $\Gamma_5$ triplet
ground state.
Here we pick up several results for $V=0.6$, $0.7$,
$0.805$, $0.8339$, and $0.9$.

For $V=0.6$, we observe a clear plateau of entropy
with the value near $\log 3$,
corresponding to the local $\Gamma_5$ triplet.
At low temperatures, the entropy $\log 3$ is eventually released,
suggesting the Kondo effect due to the screening of $S=1$,
where $S$ denotes the effective local impurity spin.
Thus, this is called the local triplet phase,
but it is considered as the Kondo singlet phase.
Readers may consider that the overscreening should occur,
but for its occurrence, relatively large values of $V_7$ and
$V_8$ are required.
In fact, as we will see later, the local triplet phase
characterized by the Kondo screening of local triplet is
widely observed in the region of small $V_7$ and $V_8$.

Next we discuss the results for $V=0.7$ and $0.805$.
Here we encounter peculiar overscreening phenomena,
where a residual entropy of $\log \phi$ is clearly observed
at low temperatures with the golden ratio $\phi=(1+\sqrt{5})/2$.
The analytic value of the residual entropy $S_{\rm ana}$
for the multi-channel Kondo effect has been given by~\cite{Affleck}
\begin{equation}
\label{eq:Sana}
S_{\rm ana}=\log \frac{\sin [(2S +1)\pi/(n_{\rm c}+2)]}
{\sin [\pi/(n_{\rm c}+2)]},
\end{equation}
where $S$ indicates the local impurity spin and
$n_{\rm c}$ denotes the number of channels.
In the present case with $n_{\rm c}=3$,
$S_{\rm ana}=\log \phi$ is easily obtained
for both the cases of $S=1/2$ and $1$.
As we will see later, it is possible to determine $S=1$
from the analysis of the quantum critical behavior between
the three-channel Kondo and Fermi-liquid phases.\cite{Hotta-Ho}

Now we turn our attention to the case of $V=0.9$
by skipping the result for $V=0.8339$.
For $V=0.9$, we observe the rapid decrease of the entropy,
suggesting the appearance of the local singlet phase.
When we change the value of $V$ from $0.8$ to $0.9$,
it is expected that a QCP appears between the three-channel
Kondo and local singlet phases.
It has been recognized that the QCP appears at the transition
between the screened Kondo and local singlet phases,
characterized by the residual entropy of $0.5 \log 2$.
\cite{Koga1,Koga2,Kusunose1,Kusunose2,OSakai,Koga3,Koga4,Miyake1,
Fabrizio1,Fabrizio2,Miyake2,Koga5,Miyake3,Miyake4,Sela,
Shiina1,Shiina2,Hotta2018,Koga6,Hotta2020,Hotta2022}
The present author has clarified that the QCP between
the two-channel Kondo and local singlet phases
is characterized by $\log \phi$.\cite{Hotta2020}

Therefore, the QCP between the three-channel Kondo and
local singlet phases is expected to be characterized by
the residual entropy of the four-channel Kondo effect.
From eq.~(\ref{eq:Sana}), for the case of $n_{\rm c}=4$,
we obtain $S_{\rm ana}=0.5 \log 3$ and $\log 2$
for $S=1/2$ and $1$, respectively.
In the $f$-electron entropies for $V=0.8339$ in Fig.~2(a),
we observe the entropy plateau with the value
between $0.5 \log 3$ and $\log 2$,
and the plateau eventually ends at around $T \sim 10^{-9}$.
This behavior is believed to denote the QCP characterized by
the residual entropy of the four-channel Kondo effect with $S=1$.

In Fig.~2(b), we show the residual entropies
at $T=1.7 \times 10^{-7}$ as a function of $x$ for $V=0.7$.
We observe the three-channel Kondo phase characterized by
the residual entropy of $\log \phi$ for a wide range of $x$
as $-0.65 < x <0.68$, corresponding to the region of
the $\Gamma_5$ triplet ground state in Fig.~1.
On the other hand, we find zero entropies for
$0.68 < x \le 1$ and $-1 \le x < -0.65$.
Since the region of $0.68 < x \le 1$ corresponds to the
$\Gamma_1$ singlet state in Fig.~1,
it is easy to understand the appearance of the local singlet phase.
The region of $-1 \le x < -0.65$ is considered to
correspond to the quasi-quintet state in Fig.~1,
but the numerical results suggest the appearance of
the local singlet phase in this region.

Let us turn our attention to the boundary region
between the three-channel Kondo and local singlet phases.
We observe sharp peaks at around $x \approx 0.685$
and $x \approx -0.646$ and the peaks assume
the values between $0.5 \log 3$ and $\log 2$.
As we have mentioned in Fig.~2(a),  this value is apparently
larger than $0.5 \log 3$, suggesting that the peak should
denote the QCP characterized by the residual entropy of
the four-channel Kondo effect with $S=1$.

Finally, we comment on the $\Gamma_5$ triplet of $n=10$.
If the existence of the $\Gamma_5$ triplet is the key condition
for the emergence of the three-channel Kondo effect,
it should often occurs in the $\Gamma_5$ triplets for
$n=2$, $4$, $8$, and $12$.
Then, we performed the NRG calculations for those cases,
but we did not find any signals of the three-channel Kondo
effect except for the case of $n=10$.
Thus, the $\Gamma_5$ triplet for $n=10$ is considered
to be special.
As shown in Fig.~2(c), \cite{Hotta-Ho}
the main components of the $\Gamma_5$ triplet for $n_b=4$
are expressed by the combination of the pseudo-spin triplet
and singlet pairs.
Since each orbital is occupied by one $f$ electron,
the $\Gamma_5$ state composed of three types of triplets
is characterized by the orbital degrees of freedom,
$\alpha$, $\beta$, and $\gamma$.
This structure of the $\Gamma_5$ triplet is important for the
occurrence of the three-channel Kondo effect.

\subsection{Results for the case of $V_7 \ne V_8$}

\subsubsection{Phase diagram and energy spectra}
\label{3-2-1}

Now we move onto the present results for the general case of
$V_7 \ne V_8$.
First, to summarize the results, we show the phase diagram
on the $(V_8, V_7)$ plane in Fig.~3(a),
including the local triplet, the local singlet, and
the three-channel Kondo phases,
which have been already suggested in Fig.~2(a).
To understand easily the correspondence with Fig.~2(a),
we draw the dotted line from the origin to $V_8=V_7=1.0$
in Fig.~3(a).
It is found that the local triplet (Kondo singlet) phase
widely spreads in the left-hand side of the phase diagram,
while the local singlet phase is basically found in the right-hand side.
Between those two phases, we find the three-channel Kondo phase
apart from the line of $V_7=0$.

\begin{figure}[t]
\centering
\includegraphics[width=8truecm]{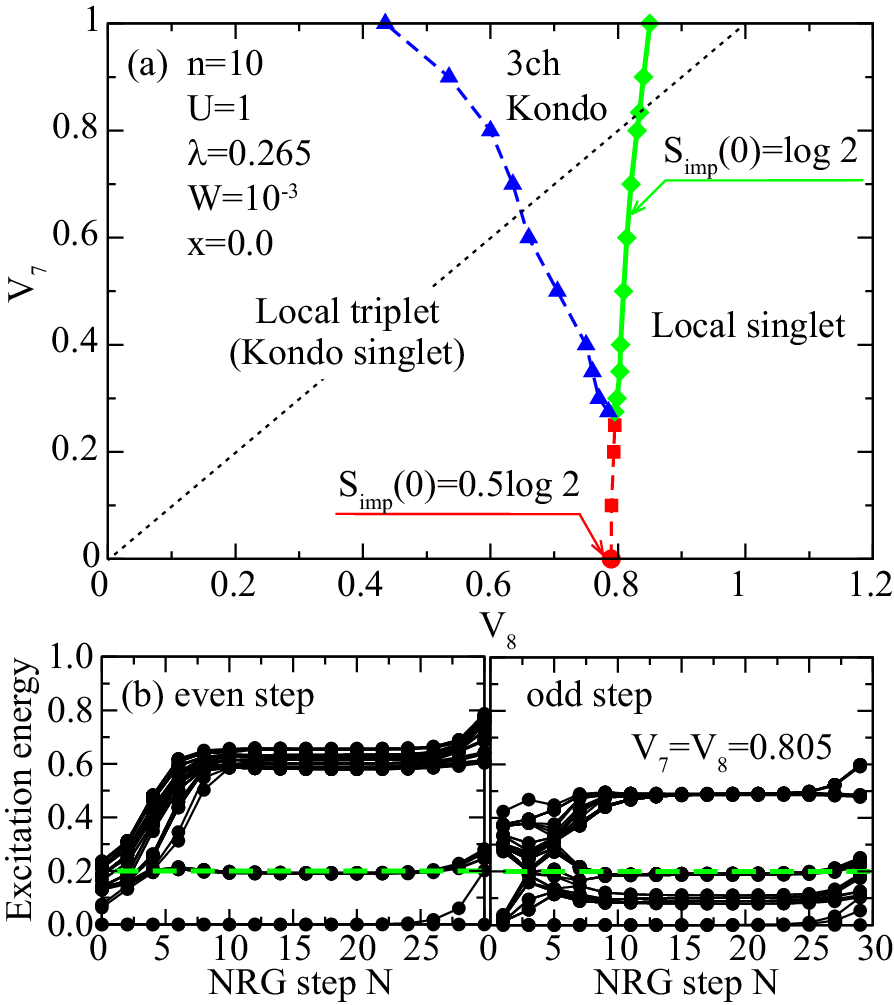}
\caption{(Color online)
(a) Phase diagram on the $(V_8, V_7)$ plane
for $0\le V_7 \le 1$ and $0 \le V_8 \le 1.2$.
The broken curves denote the boundaries determined only
by the changes in the excitation spectra,
while the solid curve indicates the boundary also characterized
by the residual entropy.
(b) Excitation energies vs. NRG steps $N$ for $V_7=V_8=0.805$
with even $N$ (left panel) and odd $N$ (right panel).
Green broken line indicates the excitation energy of $0.2$,
obtained by the conformal field theory
for the three-channel Kondo effect.\cite{Affleck,Cox2,Fabrizio}
}
\end{figure}

On the line of $V_7=0.0$, the QCP is found at the transition
between the local triplet (Kondo singlet) and local singlet phases,
characterized by the residual entropy of $0.5 \log 2$.
\cite{Koga1,Koga2,Kusunose1,Kusunose2,OSakai,Koga3,Koga4,Miyake1,
Fabrizio1,Fabrizio2,Miyake2,Koga5,Miyake3,Miyake4,Sela,
Shiina1,Shiina2,Hotta2018,Koga6,Hotta2020,Hotta2022}
However, for $0< V_7 <0.25$, we have not observed the residual
entropy at the boundary between the local triplet (Kondo singlet)
and local singlet phases, although the boundary is clearly determined
by the change in the excitation spectra, as we will show later.
For $V_7 >0.25$, the three-channel Kondo phase appears
between the local triplet (Kondo singlet) and local singlet phases.
As mentioned above, the boundary between the three-channel Kondo
and local singlet phases is characterized by the residual entropy
of $\log 2$.

Here we briefly comment on the local singlet phase,
in which the local singlet is effectively formed among $f$ electrons,
while the conduction bands are virtually separated from the impurity site.
Let us discuss the destination of the local singlet phase
when we further increase the value of $V_8$ over beyond $V_8=1.2$.
If we consider the two-orbital Anderson model,
the local singlet phase is always stabilized for large hybridization.
The present results for $V_7=0$ are essentially the same as those
of the two-orbital Anderson model.
Since the local singlet phase for $V_7=0$ is smoothly connected
to that for $V_7>0$, the local singlet phase is widely found
in the right-hand side of Fig.~3(b).
Note that for $V_7>1.0$, the situation is changed,
but this point will be discussed later.

In Fig.~3(b), we show the typical results of the excitation energies
as functions of NRG steps $N$ for $V_7=V_8=0.805$.
The left and right panels denote the results for the even
and odd $N$, respectively.
Corresponding to the entropy plateau of $\log \phi$ in Fig.~2(a),
we observe the excitation energy with the value near $0.2$,
which has been predicted by the conformal field theory
for the three-channel Kondo effect.\cite{Affleck,Cox2}
Note that in the region of $N > 25$, the deviation of the excitation
energy from $0.2$ becomes significant,
mainly due to the accumulation of numerical calculation errors.

\begin{figure}[t]
\centering
\includegraphics[width=8truecm]{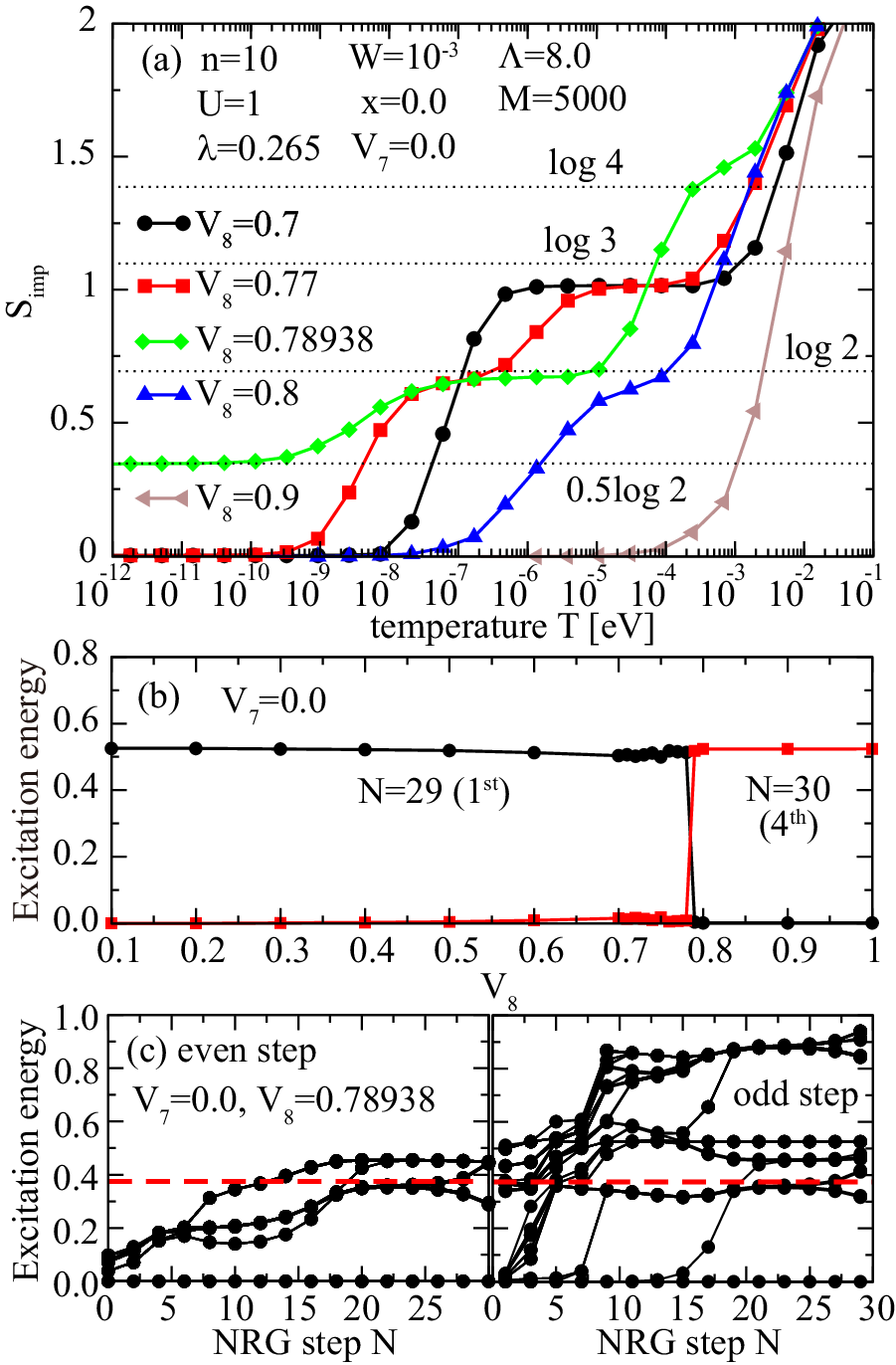}
\caption{(Color online)
(a) Entropies vs. temperature for several values of $V_8$
for $V_7=0.0$.
(b) The first excited-state energy vs. $V_8$ at $N=29$ step
and the fourth excited-state energy vs. $V_8$ at $N=30$ step
for $V_7=0.0$.
(c) Excitation energies vs. NRG steps $N$ for $V_7=0.0$ and
$V_8=0.78938$ with even $N$ (left panel) and odd $N$ (right panel).
Red broken line indicates the excitation energy of $0.375$,
which has been obtained by the conformal field theory
for the unstable fixed point of the two-channel Kondo effect.
\cite{Affleck,Fabrizio2}
}
\end{figure}

\subsubsection{Results for $V_7=0.0$}

Now we explain the NRG results in detail.
First let us discuss the results for $V_7=0.0$,
in particular, those around at the QCP.
In Fig.~4(a), we show the entropies with $V_8=0.7$, $0.77$,
$0.78938$, $0.8$, and $0.9$ for $V_7=0.0$.
For $V_8=0.7$ and $0.9$, we again find the typical behavior
for the local triplet and local singlet phases, respectively.
For $V_8=0.77$, we observe the plateau of $\log 2$
after the local triplet signal of $\log 3$, and the entropy
of $\log 2$ is eventually released at low temperatures.
For $V_8=0.8$, we observe the shoulder-like behavior of $\log 2$,
but it immediately disappears as we decrease the temperature.
Finally, for $V_8=0.78938$, we observe the plateau of $\log 2$
after the shoulder-like behavior at high temperatures.
The plateau of $\log 2$ is smoothly changed to that of
$0.5\log 2$ at low temperatures, suggesting the QCP
between the local triplet (Kondo singlet) and local singlet phases.

As shown in Fig.~4(b), the QCP at $V_8=0.78938$ corresponds
to the point at which the excitation spectra at $N=29$ and
$N=30$ are interchanged between the local triplet (Kondo singlet)
and local singlet phases.
In the local singlet phase, the electron degrees of freedom
should be suppressed at an impurity site.
Namely, the local electrons do not have any influence on
the conduction electron state.
Thus, in the local singlet phase, we expect the same energy spectrum
as that for the case with only the conduction electrons.
On the other hand, in the local triplet (Kondo singlet) phase,
the local moment of electron is screened by those of conduction
electrons.
Typically, the conduction electrons at the site next to the impurity
site form the singlet state with the local electrons.
Thus, the energy spectrum eventually becomes the same as that of
the conduction electron in the limit of large NRG step $N$,
but one step should be shifted in the energy spectra
due to the screening by conduction electrons.
Namely, the excitation spectra for even $N$ and odd $N$ are
interchanged just between the local triplet (Kondo singlet)
and local singlet phases.
Note that even for $0< V_7 <0.25$, the change in the excitation spectra
still continues to characterize the phase boundary between the local triplet
(Kondo singlet) and local singlet phases.

In Fig.~4(c), we show the excitation energies as functions of
NRG steps $N$ for $V_7=0.0$ and $V_8=0.78938$.
The left and right panels denote the results for even $N$
and odd $N$, respectively.
Corresponding to the entropy plateau of $0.5 \log 2$ in Fig.~4(a),
we observe the excitation energy near the value of $0.375$,
which has been predicted by the conformal field theory
for the unstable fixed point of the two-channel Kondo ffect.
\cite{Affleck,Fabrizio2}
This result suggests that the entropy plateau of $0.5 \log 2$
should be the signal of the QCP between the local triplet
(Kondo singlet) and local singlet phases.

\begin{figure}[t]
\centering
\includegraphics[width=8truecm]{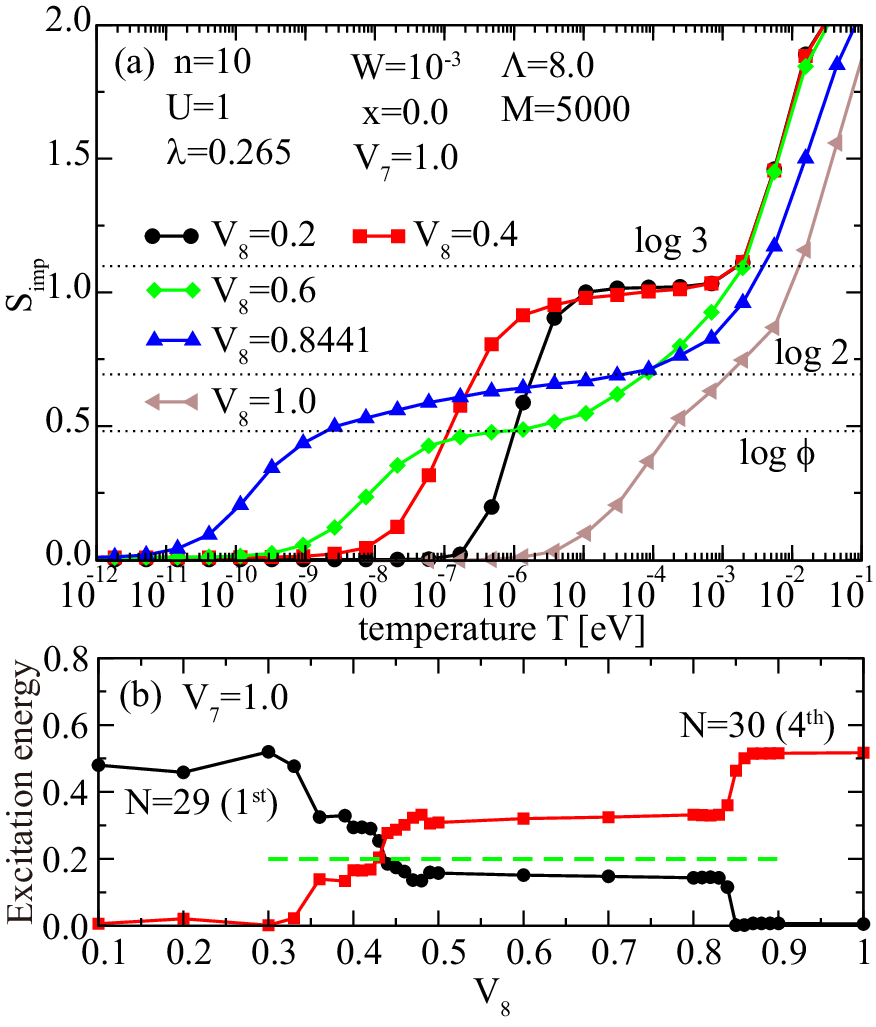}
\caption{(Color online)
(a) Entropies vs. temperature for several values of $V_8$
for $V_7=1.0$.
(b) The first excited-state energy vs. $V_8$ at $N=29$ step and
the fourth excited-state energy vs. $V_8$ at $N=30$ step
for $V_7=1.0$.
Green broken line denotes the excitation energy $0.2$,
which has been predicted by the conformal field theory
for the three-channel Kondo effect.\cite{Affleck,Cox2,Fabrizio}
}
\end{figure}

\subsubsection{Results for $V_7=1.0$}

For the case of $V_7 >0.25$,
we observe the three-channel Kondo phase
between the local triplet (Kondo singlet) and local singlet phases.
Typical results are shown in Fig.~5(a),
in which we depict entropies vs. temperature for several values
of $V_8$ for $V_7=1.0$.
For $V_8=0.2$ and $0.4$, we observe the entropy behavior
for the local triplet phase, while for $V_8=1.0$,
the local singlet phase is suggested from the entropy behavior.
Between them, for $V_8=0.6$, we find the plateau of $\log \phi$,
denoting the three-channel Kondo phase, although the length
of the plateau is limited around at $T=10^{-7} \sim 10^{-6}$.
Such entropy behavior is not persuasive to confirm
the three-channel Kondo phase.
For the confirmation, it is necessary to examine the excitation spectra,
but this point will be discussed below.
Furthermore, for $V_8=0.8441$, we again encounter
the remnant of a residual entropy of $\log 2$,
suggesting the QCP between the three-channel Kondo
and local singlet phases.
Namely, the boundary between the three-channel Kondo
and local singlet phases is defined by the quantum critical behavior
such as the appearance of the residual entropy of $\log 2$.

However, the boundary between the local triplet (Kondo singlet) and
the three-channel Kondo phases is not characterized by the residual
entropy behavior.
To find the boundary between the local triplet (Kondo singlet) and
the three-channel Kondo phases, we investigate the excitation spectra.
In Fig.~5(b), we show the first excited-state energy of $N=29$ and
the fourth excited-state energy of $N=30$ as functions of $V_8$
for $V_7=1.0$ along the upper edge of the phase diagram in Fig.~3(a).
When we compare the excitation energies of the local triplet
(Kondo singlet) and local singlet phases, it is possible to observe
the same structure as in Fig.~4(b).
However, in the three-channel Kondo phase, we find the excitation
spectra different both from those in the local triplet (Kondo singlet) and
local singlet phases.

Namely, the first excited-state energy seems to take the
value near $0.2$, as pointed out in Fig.~3(b).
The value of $0.2$ has been analytically obtained
for the three-channel Kondo phase by the conformal field theory.
\cite{Affleck,Cox2,Fabrizio}
The calculated value is obviously deviated from $0.2$,
but it is different from that of the Fermi-liquid phase.
Thus, this behavior is considered to be a signal of
the three-channel Kondo phase.
The deviation from the analytic value is considered to be due to
the precision of the numerical calculations,
indicating that the value should approach the analytic value
when we increase the number of $M$ and decrease
the cut-off $\Lambda$.
For $V_8$ in the range of $0.3 < V_8 <0.5$,
when we increase $V_8$, the excitation energies for $N=29$
and $N=30$ gradually decrease and increase, respectively,
leading to the interchange between them around at $V_8 \approx 0.43$.
This value is considered to define the boundary between the local triplet
(Kondo singlet) and the three-channel Kondo phases.

\begin{figure}[t]
\centering
\includegraphics[width=8truecm]{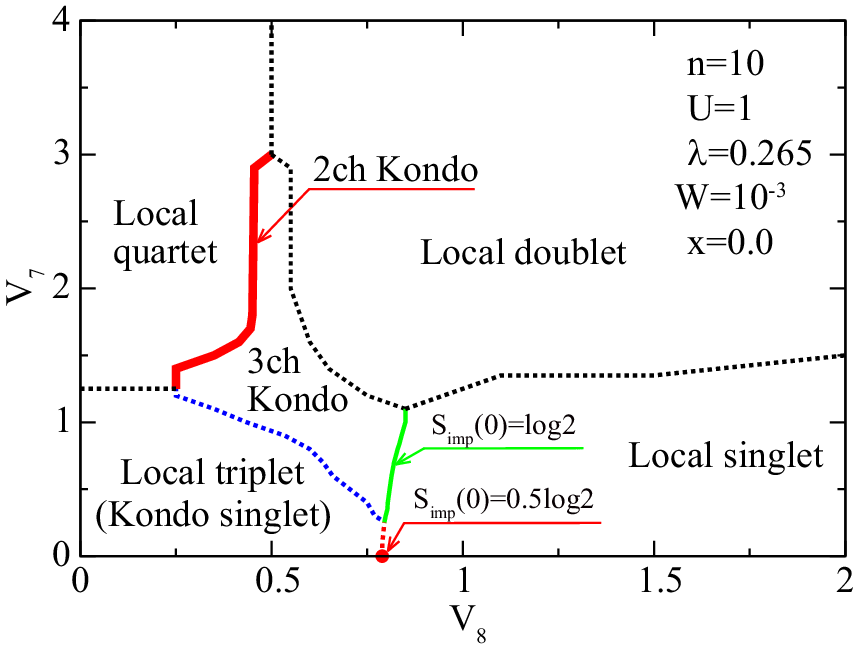}
\caption{(Color online)
Phase diagram on the $(V_8, V_7)$ plane
for $0\le V_7 \le 4$ and $0 \le V_8 \le 2$.
The results in the region of  $0\le V_7 \le 1$ and $0 \le V_8 \le 1.2$
are the same as those in Fig.~3(b).
Note that the broken curves denote the boundaries
determined only by the changes in the excitation spectra,
while the solid curves indicates the boundaries characterized
both by the residual entropy and the changes in the excitation spectra.
The meaning of the thick solid curve is explained in the maintext.
}
\end{figure}

\subsubsection{Phase diagram in the wide parameter space}

From the results of residual entropies and excitation spectra,
we have confirmed the existence of the three-channel Kondo phase
between the local triplet (Kondo singlet) and local singlet phases.
Here readers may have a naive question about the destination
of the three-channel Kondo phase,
when we increase the value of $V_7$ over beyond $V_7=1.0$.
To answer this point, it is necessary to expand the phase diagram
outside the range of $(V_8, V_7)$ in Fig.~3(a).

Figure 6 indicates the phase diagram in the region of
$0  \le V_7 \le 4$ and $0 \le V_8 \le 2$.
Here we honestly mention that the boundary curves for $V_7 >1$
are not smoothly depicted in comparison with Fig.~3(a),
since it was hard tasks to collect enough numerical data so as
to depict all the boundaries in the same precision as in Fig.~3(a).
However, we believe that the essential points can be grasped
in the present figure.

In Fig.~6, we observe a couple of new phases
as local quartet and local doublet phases,
which have not been found in Fig.~3(a).
Note that they are considered to be Fermi-liquid phases.
Later we will discuss in detail the entropy behavior in
these two phases.
As for the destination of the three-channel Kondo phase
when we increase the value of $V_7$,
it is not difficult to imagine the tendency that
the three-channel Kondo phase is eventually closed
in the range of $V_7>1$.
However, it is a surprising issue that the three-channel Kondo
phase still survives with a narrow region along the line of
$V_8=0.5$ up to $V_7=3.0$.
This point will be discussed later.

Here we provide a comment on the local doublet phase in Fig.~6.
As emphasized in Sect.~\ref{3-2-1},
the local singlet phase on the line of $V_7=0$ is
smoothly connected to that for $0<V_7<1.0$.
Then, we have concluded that the local singlet phase appears
even for large $V_8$ in the region of $0 < V_7<1.0$.
However, in the region of large $V_8$ for $V_7 > 1.0$,
we consider another possibility that the Kondo-like phase
occurs instead of the local singlet phase.
This is just the local doublet phase in Fig.~6, which is
stabilized to gain the effect of $V_8$
with the assistance of $V_7$.
A way to distinguish the local singlet and doublet phases
will be discussed in Sect.~\ref{3-2-8}.

\begin{figure}[t]
\centering
\includegraphics[width=8truecm]{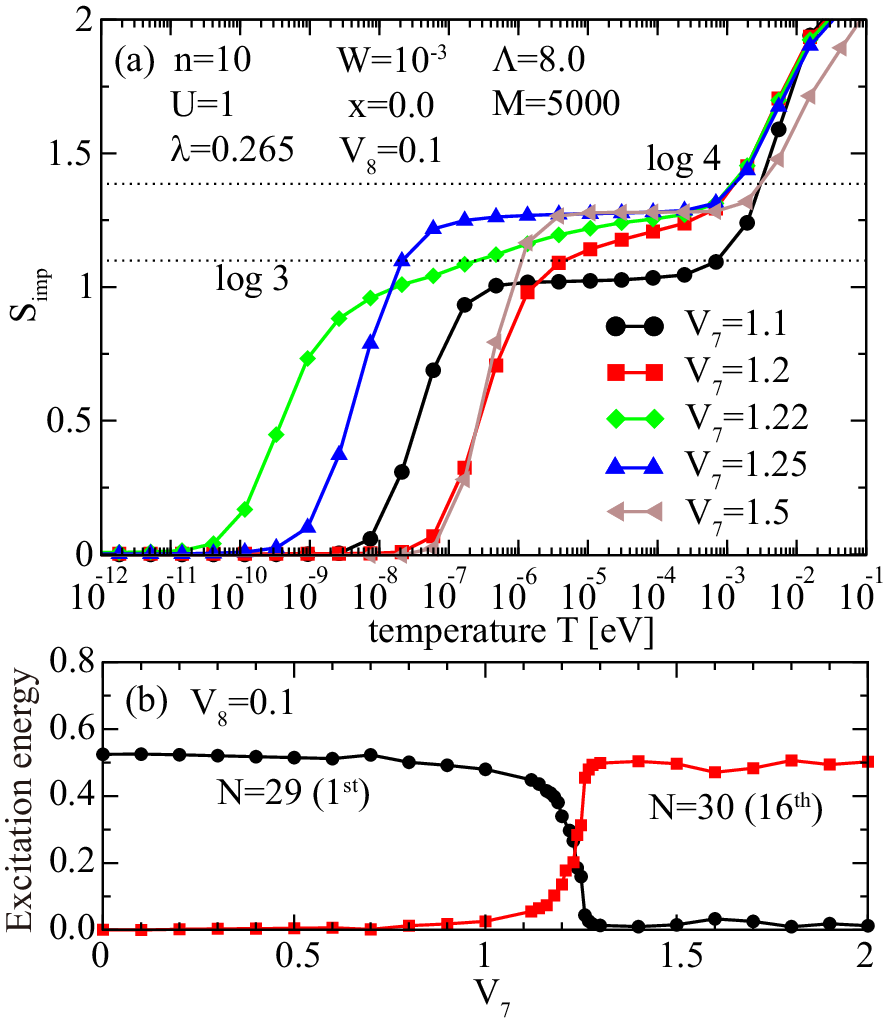}
\caption{(Color online)
(a) Entropies vs. temperature for several values of $V_7$ for $V_8=0.1$.
(b) The first excited-state energy vs. $V_7$ at $N=29$ step and
the sixteenth excited-state energy vs. $V_7$ at $N=30$ step
for $V_8=0.1$.
}
\end{figure}

\subsubsection{Results for $V_8=0.1$}

Now let us discuss the changes in the entropy and the excitation
energy along the line of $V_8=0.1$.
In Fig.~7(a), we show the NRG results of $f$-electron entropy
for $W=10^{-3}$, $x=0.0$, and $V_8=0.1$
with the several values of $V_7$ in the range of $1.1 \le V_7 \le 1.5$
across the boundary between the local triplet and quartet phases.
For $V_7=1.1$, we observe an entropy plateau with the value
near $\log 3$ and it is eventually released to move to the singlet state.
This is the same behavior as mentioned in the local triplet phase,
which we have found for small $V_7$ and $V_8$ in Fig.~3(a).

When we increase the value of $V_7$, we encounter
the different behavior.
Namely, for $V_7=1.25$ and $1.5$, an entropy plateau
with the value near $\log 4$ is clearly observe and
it eventually disappears at low temperatures.
Thus, this phase is called the local quartet, but the appearance
of the local quartet is easily understood as follows.
Let us consider the limiting case of $V_8=0$.
For large $V_7$, the local $\Gamma_7$ electron is strongly hybridized
with $\Gamma_7$ conduction electron and the remaining
three $f$ electrons, one $\Gamma_6$ and two $\Gamma_8$,
are considered to form the local $\Gamma_8$ quartet.

Between the local triplet and quartet phases,
the change in the entropy plateau does not occur abruptly.
In Fig.~7(a), for $V_7=1.2$, $1.22$, and $1.25$,
the values of the entropy plateaus are changed gradually
from $\log 3$ to $\log 4$.
To determine the boundary between the local triplet and local
quartet phases, it is useful to investigate the excitation spectra.
In Fig.~7(b), we show the first excited-state energy of $N=29$
and the sixteenth excited-state energy of $N=30$
as functions of $V_7$ for $V_8=0.1$.
We observe that a couple of excitation energies are interchanged
around at $V_7=1.22$, suggesting the boundary
between the local triplet and local quartet phases.
By tracking the boundaries when we change the value of $V_8$,
we depict the boundary line for small $V_8$ in Fig.~6.

\begin{figure}[t]
\centering
\includegraphics[width=8truecm]{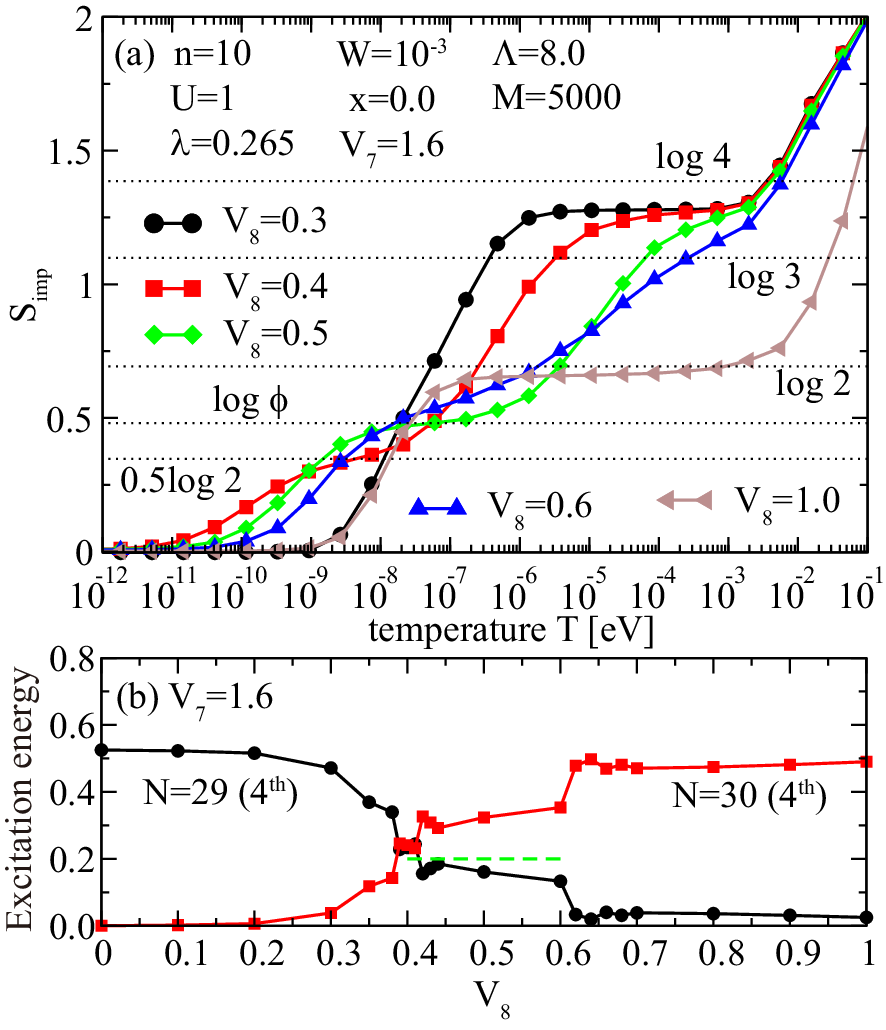}
\caption{(Color online)
(a) Entropies vs. temperature for several values of $V_8$
for $V_7=1.6$.
(b) The fourth excited-state energy vs. $V_8$ at $N=29$ step and
the fourth excited-state energy vs. $V_8$ at $N=30$ step
for $V_7=1.6$.
Green broken line indicates the excitation energy of $0.2$,
predicted by the conformal field theory
for the three-channel Kondo effect.\cite{Affleck,Cox2,Fabrizio}
}
\end{figure}

\subsubsection{Results for $V_7=1.6$}

Now we turn our attention to the destination of the
three channel Kondo phase when we increase the value of $V_7$.
In Fig.~8(a), we show the NRG results of $f$-electron entropy
for $W=10^{-3}$, $x=0.0$, and $V_7=1.6$
with the several values of $V_8$ in the range of $0 \le V_8 \le 1.0$.
First we remark the result for $V_8=1.0$, in which we encounter
the plateau with the value near $\log 2$ at high temperatures,
but it is eventually released at low temperatures.
Thus, it is called the local doublet phase.
Intuitively, the appearance of the local doublet is considered to
originate from the localized $\Gamma_6$ electron,
since the $\Gamma_7$ and $\Gamma_8$ electrons are
dragged out by the three conduction bands
for relatively large values of both $V_7$ and $V_8$.
Note that the local doublet phase is considered to be Fermi liquid.

Next we turn our attention to the cases for small $V_8$.
For $V_8=0.3$, we find the entropy plateau with the value
near $\log 4$ as observed in the local quartet phase.
For $V_8=0.5$, the entropy plateau with the value of $\log \phi$
can be observed and it is the signal of the three-channel Kondo phase,
as mentioned before.
For $V_8=0.6$, we observe the remnant of the plateau of $\log \phi$,
but it is also considered to suggest the existence of
the three-channel Kondo phase.

Here we remark the entropy behavior for $V_8=0.4$.
In this case, we also observe the remnant of the plateau,
but the value denotes $0.5 \log 2$, not $\log \phi$.
This is considered to be the signal of QCP or the existence
of the two-channel Kondo phase.
To clarify this point, it is highly recommended to check
the excitation spectra.

For the purpose, in Fig.~8(b), we show the fourth excited-state
energy of $N=29$ and the fourth excited-state energy of $N=30$
as functions of $V_8$ for $V_7=1.6$.
For $V_8 < 0.4$ and $V_8 > 0.6$, we find the behavior
of the local quartet and singlet phases, respectively.
For $0.42 \le V_8 \le 0.6$, we observe the excitation energy
near $0.2$ at $N=29$ steps, suggesting the three-channel
Kondo phase.\cite{Affleck,Cox2,Fabrizio}
Here we show only the values at $N=29$ and $30$,
but we could obtain the $N$ dependence of the
excitation spectra, similar to those in Fig.~3(b).

Note that we find peculiar behavior in the narrow range of
$0.39 \le V_8 \le 0.41$ in Fig.~8(b).
This is considered to the signal of the two-channel Kondo phase,
although the values of the excitation energies at $N=29$ and $30$
are deviated from that predicted by the conformal field theory.
As for this value, we will discuss it later, but here we provide
a couple of comments from a qualitative viewpoint.
First, if this is the QCP, the excitation energies should be interchanged
at the critical value, as already shown in Fig.~4(b).
However, we observe the finite range of $0.39 \le V_8 \le 0.41$
in the excitation energies.
We deduce that it is the signal for the appearance of
the two-channel Kondo phase, not QCP.
Second, to express the finite range for the two-channel Kondo phase
between the local quartet and the three-channel Kondo phases,
we depict the boundary curve between those two phases
by the thick solid line in Fig.~6.

\begin{figure}[t]
\centering
\includegraphics[width=8truecm]{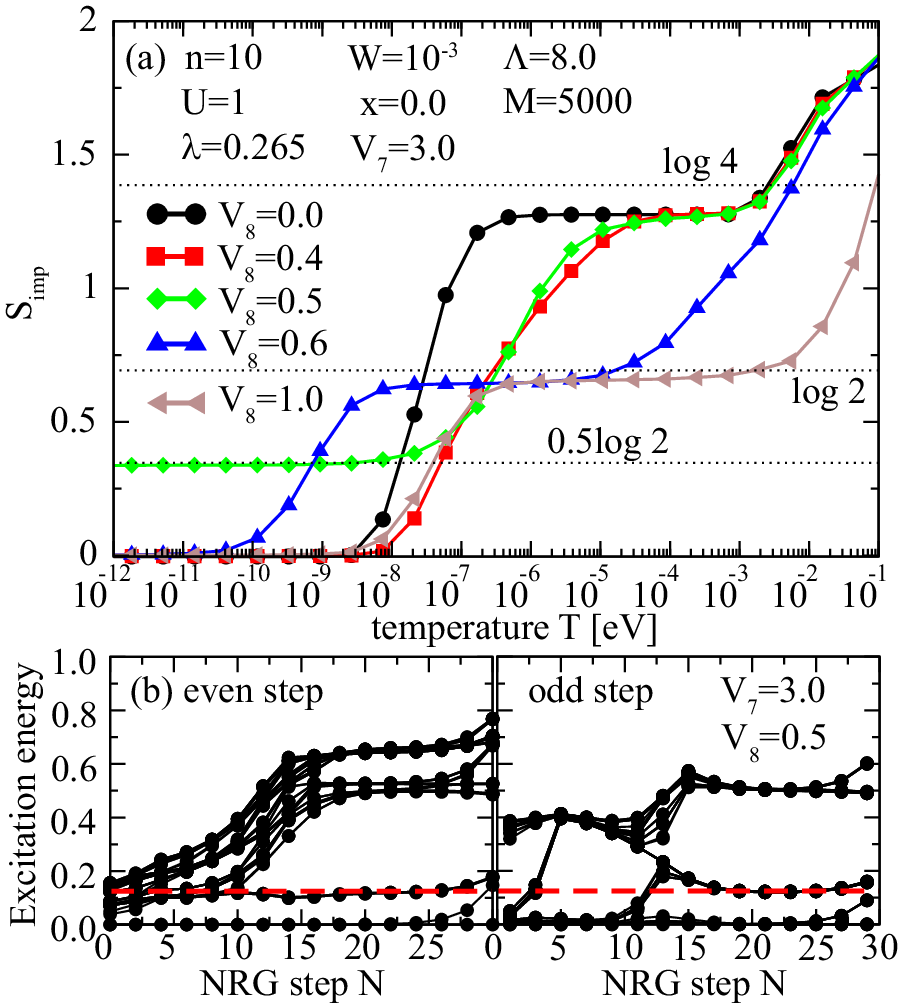}
\caption{(Color online)
(a) Entropies vs. temperature for several values of $V_8$ for $V_7=3.0$.
(b) Excitation energies vs. NRG steps $N$ for $V_8=0.5$ and
$V_7=3.0$ with even $N$ (left panel) and odd $N$ (right panel).
Red broken line indicates the excitation energy of $0.125$,
predicted by the conformal field theory
for the stable fixed point of the two-channel Kondo effect.
\cite{Affleck,Cox2}
}
\end{figure}

\subsubsection{Results for $V_7=3.0$}

To promote our understanding on the two-channel Kondo phase
appearing in the region of small $V_8$,
in Fig.~9(a), we show the NRG results of $f$-electron entropy
for $W=10^{-3}$, $x=0.0$, and $V_7=3.0$
with the several values of $V_8$ in the range of $0 \le V_8 \le 1.0$.
For $V_8=0.0$ and $0.4$, we find the plateau with the value near
$\log 4$ which is eventually released at low temperatures,
indicating the signal of the local quartet phase.
On the other hand, for $V_8=0.6$ and $1.0$,
we encounter the plateau with the value near $\log 2$
which is also released at low temperatures.
This behavior is considered to be the signal of
the local doublet phase.

Let us here concentrate on the case of $V_8=0.5$,
in which we find the plateau with the value near $\log 4$
at high temperatures.
However, when we decrease the temperature,
we clearly observe another plateau with the value of $0.5 \log 2$.
This behavior is considered to be the signal of QCP or
the existence of the two-channel Kondo phase.
To clarify this point, we plot the excitation energies as
functions of NRG steps $N$ for $V_7=3.0$ and $V_8=0.5$
in Fig.~9(b).
The left and right panels denote the results for even $N$
and odd $N$, respectively.

Corresponding to the entropy plateau of $0.5 \log 2$
in Fig.~9(a) for $V_8=0.5$, we observe the excitation
energy of $0.125$, not $0.375$, which has been predicted
by the conformal field theory for the stable fixed point
of the two-channel Kondo effect.\cite{Affleck,Cox2}
Thus, the entropy plateau of $0.5 \log 2$
at $(V_8, V_7)=(0.5, 3.0)$
suggests the two-channel Kondo phase, not the QCP.
Since this is the same conclusion as that in the region of
$0.39 \le V_8 \le 0.41$ for $V_7=1.6$,
the boundary between the local quartet and the three-channel
Kondo phases is depicted by the thick red line in Fig.~6.
Note that in this paper, we do not show the finite range of
$V_8$ for the two-channel Kondo phase and the details of
the edge of the two-channel Kondo phase.
For the purpose to clarify these points, more precise calculations
are required, but they are postponed as one of the future tasks.

\subsubsection{Boundary between local singlet and doublet phases}
\label{3-2-8}

The figure about the boundary between the local singlet and doublet
phases is not shown here, since it has no direct relation with
the three-channel Kondo phase,
but we provide a brief comment on this boundary.
When we consider the region far from the boundary,
it is easy to distinguish them only from the entropy behavior.
Namely, for the local doublet phase, first we find the
entropy plateau of $\log2$ and it eventually released
at low temperatures.
On the other hand, for the local singlet phase,
the entropy rapidly becomes zero even at high temperatures
in the order of $0.1$.

However, in the vicinity of the boundary, the temperature
dependences of the entropy of those two phases are similar
to each other.
To distinguish them, it is necessary to investigate the change
in the excitation spectra.
When we plot the sixteenth excited-state energy of $N=29$
and the first excited-state energy of $N=30$
as functions of $V_7$ for the fixed value of $V_8$
in the region of $0.8 \le V_8 \le 2.0$,
it is found that two excitation energies are interchanged
at a certain value of $V_7$, leading to the boundary
between the local singlet and doublet phases.
By repeating the NRG calculations, we could depict
the boundary curve in Fig.~6.

\section{Discussion and Summary}

In this paper, we have investigated the three-channel Kondo phase
appearing for the case of Ho$^{3+}$ ion with ten $4f$ electrons
by analyzing numerically the seven-orbital impurity Anderson model
hybridized with $\Gamma_7$ and $\Gamma_8$ conduction electron bands.
From the residual entropy of $\log \phi$ and the excitation energy spectra,
we have confirmed the emergence of the three-channel Kondo effect
for the local $\Gamma_5$ triplet ground state.
We have also found the three-channel Kondo phase in a wide range
on the $(V_8, V_7)$ plane, surrounded by Fermi-liquid phases
such as local singlet, doublet, triplet, and quartet phases.
The boundary curves among them have been determined by
the entropy behavior and the change in the excitation energy spectra.

Among the boundary curves in the phase diagram Fig.~6,
it is necessary to mention honestly the red thick lines
indicating the two-channel Kondo phase,
found in the region of $1.2<V_7<3.0$ and $0.25<V_8<0.5$.
As mentioned in Fig.~8(b), at least for $V_7=1.6$,
we have found the two-channel Kondo phase
in the narrow range of $0.39 \le V_8 \le 0.41$.
To express the narrow range, we have used the thick line
in the phase diagram, but unfortunately, it may not be
correct in the exact sense.
Namely, it is necessary to depict a couple of boundary curves.
One is the boundary between the local quartet (Fermi liquid)
and the two-channel Kondo phases.
Another is the boundary between the two-channel Kondo and
the three-channel Kondo phases.
The former boundary curve is characterized by the QCP with
the residual entropy of $\log \phi$,\cite{Hotta2022}
whereas the latter one is probably related with the unknown QCP,
since it is the boundary between different non-Fermi liquid phases.
It is interesting to clarify the signal of this QCP in the entropy behavior.
However, to draw such two boundary curves, it is necessary to perform
the NRG calculations on the $(V_8, V_7)$ plane
which should be divided into much smaller meshes.
Such calculations heavily consume the CPU time and thus,
we postpone such a task in future.

In the phase diagrams, Figs.~3(a) and 6, we have found
the local singlet, doublet, triplet, and quartet phases.
Except for the local singlet phase, the Kondo temperature $T_{\rm K}$
should be defined from the screening of the local moment.
An easy guideline of $T_{\rm K}$ is the peak position of the specific heat
$C_{\rm imp}$, which is defined from the entropy $S_{\rm imp}$
as $C_{\rm imp}=T \partial S_{\rm imp}/\partial T$,
since $T_{\rm K}$ is considered to be the temperature
at which the entropy is released.
However, in the present calculations, when $T_{\rm K}$ becomes
smaller than $10^{-8}$, the magnitudes of $T_{\rm K}$
in the local doublet, triplet, and quartet phases
do not seem to depend correctly on the values of $V_7$ and $V_8$.
This is due to the problem in the precision of the present NRG calculations.
To improve this point, it is necessary to increase the value of $M$
and decrease the cut-off $\Lambda$.
Such NRG calculations need the large memory size in addition to the CPU time.
This is also a future problem.

As mentioned in Sect.~3, it is recommended to improve the precision of the
boundary curves in Fig.~6 in comparison with those in Fig.~3(a).
In particular, the boundary curves surrounding the three-channel Kondo phase
should be redrawn by more precise calculations,
although we believe that the essential points in the present phase diagram
are not changed.
To redraw the phase diagram, it is necessary to repeat the NRG calculations
in more fine meshes, but such calculations heavily consume the CPU time.
This is another future task.

Finally, we provide a short comment on the emergence of the three-channel
Kondo effect in actual materials.
Among cubic Ho compounds, HoCo$_2$Zn$_{20}$ has been recently synthesized
by the research group of Japan Atomic Energy Agency.\cite{Ho1-2-20}
Unfortunately, the signal of the three-channel Kondo effect has not been
confirmed yet, but it has been observed that the temperature dependences
of HoCo$_2$Zn$_{20}$ in the resistivity and the magnetization are
similar to those of NdCo$_2$Zn$_{20}$.\cite{Nd1-2-20}
In addition, the analysis of the $4f$ electron states at Ho site
has suggested the importance of the hyperfine interaction
between $4f$ electrons and the Ho nuclear spin.\cite{Ho1-2-20}
It is intriguing that the three-channel Kondo effect is suppressed
or not by the existence of the hyperfine interaction.
This is a challenging future problem.

In summary, we have shown the phase diagram of the
seven-orbital impurity Anderson model for the case of $n=10$
corresponding to Ho$^{3+}$ ion by performing the NRG calculations.
The phase diagram has included the three-channel Kondo phase,
surrounded by the local singlet, doublet, triplet, and quartet phases.
The boundary curves among those phases have been determined
by the entropy behavior and the excitation spectra.
We believe that the existence of the three-channel Kondo
phase for Ho$^{3+}$ ion is widely confirmed.
It is an interesting future issue to detect experimentally
the three-channel Kondo phase in Ho 1-2-20 compound.

\section*{Acknowledgment}

The author thanks Y. Haga, S. Kambe, T. Kitazawa, K. Kubo, H. Sakai,
and Y. Tokunaga for discussions and comments.
The computation in this work was partly done using the facilities of the
Supercomputer Center of Institute for Solid State Physics, University of Tokyo.



\begin{thebibliography}{99}
\bibitem{footnote}
In the previous papers, the off-diagonal CEF terms
eq.~(\ref{eq:CEFoff}) were inadvertently suppressed
in the local Hamiltonian, but such terms have been correctly
included in the code for the numerical calculation.
\end{thebibliography}
\end{document}